\documentclass[twocolumn,,amsmath,amssymb]{revtex4}
\pdfoutput=1

\usepackage{graphicx}
\usepackage{dcolumn}
\usepackage{bm}
\newcommand{\etal}{\emph{et al.}}
\newcommand{\be}{\begin{equation}}
\newcommand{\ee}{\end{equation}}
\newcommand{\bfig}{\begin{figure}}
\newcommand{\efig}{\end{figure}}

\begin{document}      
\title{Signature of the chiral anomaly in a Dirac semimetal -- a current plume steered by a magnetic field\footnote{Article based on a talk by NPO at the APS March Meeting, San Antonio, 2015.}
}  

\author{Jun Xiong$^{1}$}
\author{Satya K. Kushwaha$^{2}$}
\author{Tian Liang$^1$}
\author{Jason W. Krizan$^{2}$}
\author{Wudi Wang$^{1}$}
\author{R. J. Cava$^{2}$}
\author{N. P. Ong$^{1}$}
\affiliation{
Departments of Physics$^1$ and Chemistry$^2$, Princeton University, Princeton, NJ 08544
} 

\date{\today}      
\pacs{}
\begin{abstract}
In this talk$^*$, we describe recent experimental progress in detecting the chiral anomaly in the Dirac semimetal Na$_3$Bi in the presence of a magnetic field. The chiral anomaly, which plays a fundamental role in chiral gauge theories, was predicted to be observable in crystals by Nielsen and Ninomiya in 1983~\cite{Nielsen}. Theoretical progress in identifying and investigating Dirac and Weyl semimetals has revived strong interest in this issue~\cite{Ashvin,Balents,Ran,Son,Hosur}. In the Dirac semimetal, the breaking of time-reversal symmetry by a magnetic field $\bf B$ splits each Dirac node into two chiral Weyl nodes. If an electric field $\bf E$ is applied $\parallel \bf B$, charge is predicted to flow between the Weyl nodes. We report the observation in the Dirac semimetal Na$_3$Bi of a novel, negative and highly anisotropic magnetoresistance (MR). We show that the enhanced conductivity has the form of a narrowly defined plume that can be steered by the applied $\bf B$. The novel MR is acutely sensitive to deviations of $\bf B$ from $\bf E$, a feature incompatible with conventional transport. The locking of the current plume to $\bf B$ appears to be a defining signature of the chiral anomaly. 
\end{abstract}
 
\maketitle      
\section{Preamble}
\begin{figure}[t]
\includegraphics[width=3cm]{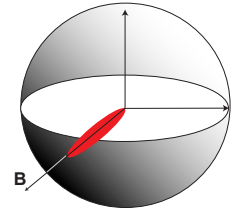}
\caption{\label{plume} 
A narrow current plume steered by magnetic field $\bf B$.
}
\end{figure}

The intriguing possibility of observing the chiral (or axial) anomaly as a novel charge current in a crystal has been discussed since 1983~\cite{Nielsen}. In the field of topological phases of matter, this question has lately received intense interest in the context of Weyl semimetals~\cite{Ashvin,Balents,Ran,Son,Sid,Hosur}. To realize a Weyl semimetal, a promising path appears to be to start with a 3D Dirac semimetal~\cite{Young,Bernevig}, and then split each Dirac node into a pair of Weyl nodes using a magnetic field $\bf B$. Rapid progress in identifying candidates for Dirac semimetals has been made~\cite{Murakami1,Murakami2,Young,Bernevig,Wang1,Wang2,Nagaosa}. The two semimetals identified~\cite{Wang1,Wang2}, Na$_3$Bi and Cd$_3$As$_2$, were recently confirmed to be Dirac semimetals by angle-resolved photoemission (ARPES)~\cite{Chen,Borisenko,Neupane,Liu,Suyang}, scanning tunneling microscopy~\cite{Yazdani} and transport experiments~\cite{Liang,Xiong1}. We focus here on the results from a recent experiment in which the axial current in Na$_3$Bi is detected as a collimated current plume locked to the direction of $\bf B$~\cite{Xiong2}. Figure \ref{plume} summarizes pictorially our main finding.

\begin{figure}[t]
\includegraphics[width=9cm]{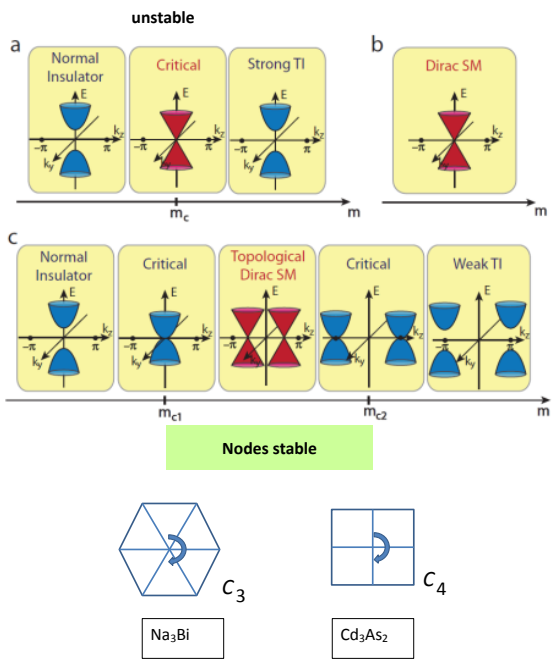}
\caption{\label{figPhase} 
Comparison of band closing scenarios as a function of the tuning parameter ${\cal P}$ (here written as $m$). The upper panel (a) shows the accidental band-closing case in which TRS and IS stabilize the Dirac node at a high-symmetry point at one critical value $m_c$. The lower panel shows the case when TRS, IS and $C_n$ protect the Dirac node on a symmetry axis over the finite interval $m_{c1}\to m_{c2}$ (reproduced from Yang and Nagaosa~\cite{Nagaosa}). The two predicted~\cite{Wang1,Wang2} Dirac semimetals Na$_3$Bi and Cd$_3$As$_2$ correspond to $C_3$ and $C_4$ point group symmetries.
}
\end{figure}

\section{Node protection by symmetry and Weyl nodes}
The problem of how the energy gap in a narrow-gap semiconductor closes and re-opens as a function of a tuning parameter ${\cal P}$ (e.g. pressure, doping or temperature) has received considerable attention in the context of topological phases and 3D Dirac states in the bulk~\cite{Murakami1,Murakami2,Young,Bernevig,Wang1,Wang2,Nagaosa}. The situation in which the gap closes at a single, critical value ${\cal P}_c$ used to be regarded as the norm (Fig. \ref{figPhase}a, b). This situation -- dubbed accidental band crossing --  is inherently unstable and not very interesting experimentally. Recent progress has shown that Dirac nodes can be protected by symmetry over an extended range of ${\cal P}$ (and hence stable) (Fig. \ref{figPhase}c). If time-reversal symmetry (TRS) and inversion symmetry (IS) are both present, the protection extends only to Dirac nodes pinned to high-symmetry points (either $\Gamma$ or the corners of the Brillouin Zone)~\cite{Young}. Unfortunately, this severely limits the number of realistic candidate materials. Later, it was realized that inclusion of point-group symmetry (PGS) extends the protection to nodes away from the high-symmetry points provided they lie on a symmetry axis~\cite{Bernevig,Wang1,Wang2,Nagaosa}. This crucial insight, which greatly expands the list of materials, led to the prediction by the IOP group~\cite{Wang1,Wang2} that Na$_3$Bi (with the PGS $C_3$) and Cd$_3$As$_2$ ($C_4$) are Dirac semimetals with protected nodes (Fig. \ref{figPhase}c). In the past year, ARPES~\cite{Chen,Borisenko,Neupane,Liu,Suyang}, STM~\cite{Yazdani} and transport experiments~\cite{Xiong1,Liang} have confirmed these predictions. Recently, surface FS arcs have been detected by ARPES~\cite{HasanFS}.

\begin{figure}[t]
\includegraphics[width=8cm]{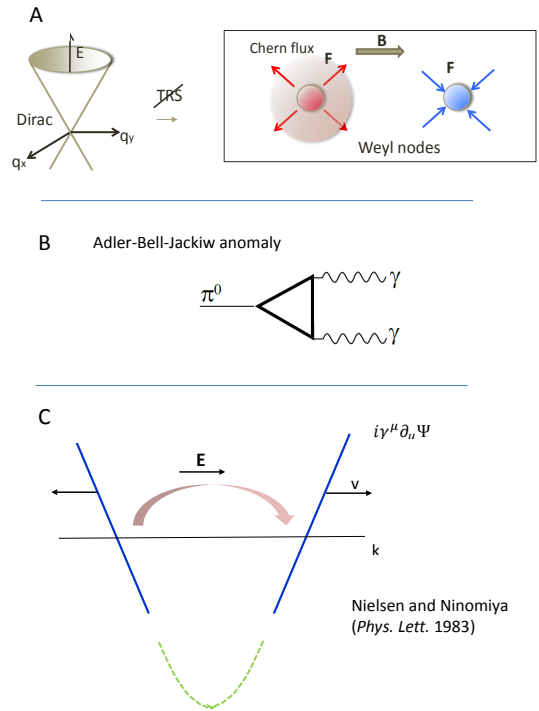}
\caption{\label{figWeyl} 
Panel A: Breaking of time reversal symmetry (TRS) in an applied $\bf B$ splits the Dirac node into a pair of Weyl nodes with opposite $\chi$. The Weyl node with $\chi = 1$ (-1) is a monopole source (sink) of Berry curvature $\vec{\cal F}$. The Chern flux captured by the Gauss surface $\cal S$ (pink sphere) normalized by 2$\pi$ gives the value of $\chi$.
Panel B: The Adler-Bell-Jackiw anomaly describing the very rapid decay of $\pi^0$ into two photons $\gamma$~\cite{Adler,Bell}.
Panel C: The proposal of Nielsen and Ninomiya~\cite{Nielsen} for observing the chiral anomaly in a 1D crystal. The slanted lines represent the dispersion of right- and left-moving chiral fermions described by the massless Dirac equation $i\gamma^\mu\partial_{\mu}\Psi = 0$. Chiral charge conservation is broken when an $E$-field is applied. Charge is pumped between the two branches. The dashed curve indicates the connection between the chiral branches deep in the Fermi sea in the solid-state context.
}
\end{figure}

When a Dirac semimetal is immersed in a strong magnetic field $\bf B$, one of the 3 protecting symmetries, TRS, is removed. Instead of reopening the gap, each of the Dirac nodes is predicted to split into two Weyl nodes~\cite{Ashvin,Balents,Ran,Son,Hosur} (Fig. \ref{figWeyl}A). The Weyl nodes, which may be regarded as monopole sources and sinks of Berry curvature $\vec{\cal F}$ in $\bf k$ space, come in pairs with opposite chiralities $\chi = \pm 1$. It is helpful to regard $\vec{\cal F}(\bf k)$ as an effective magnetic field that lives in $k$-space. If we surround a Weyl node with a Gauss surface ${\cal S}$, the Chern flux captured defines the chirality $\chi = (1/2\pi)\oint_S \vec{\cal F}({\bf k})\cdot \vec{dS}({\bf k})$. Thus, breaking TRS produces an intense Berry curvature $\vec{\cal F}$ in the vicinity of each Weyl node. Given that $\vec{\cal F}$ imparts a transverse ``anomalous'' velocity ${\bf v}_A = \vec{\cal F}\times e{\bf E}$ to a wave packet, we should expect a host of novel transport features ($\bf E$ is an applied $E$-field). The most interesting involves the chiral anomaly.

\section{Chiral anomaly}
A bit of topological physics fell into quantum field theory (QFT) in the late 1960s~\cite{Adler,Bell,Peskin,Zee}. The charged pions $\pi^{\pm}$ are remarkably long-lived mesons (lifetimes $\tau\sim 2.3\times 10^{-8}$ s) because, being the lightest hadron, they can only decay by the weak interaction into muons and neutrinos via the processes $\pi^{-} \to \mu^- + \bar{\nu}_\mu$ and $\pi^{+}\to \mu^+ + \nu_\mu$. Mysteriously, the neutral pion $\pi^0$ decays much more quickly (by a factor of 300 million) even though it is a member of the same isospin triplet. Instead of the slow leptonic channels, $\pi^0$ decays by coupling to the electromagnetic field $F_{\mu\nu}$ in the process $\pi^0\to 2\gamma$. The relevant diagram (Fig. \ref{figWeyl}B), called the Adler-Bell-Jackiw anomaly~\cite{Adler,Bell}, is a triangular fermion loop that links the $\pi^0$ (the axial current) to the 2 photons (vector currents)~\cite{Peskin,Zee}. A hint of the topological nature of the anomaly is that the one-loop diagram receives no further corrections to all orders of perturbation theory. Subsequent research revealed that the anomaly expresses the breaking of a classical symmetry by quantum fluctuations. In modern QFT, the anomaly plays the fundamental role of killing unviable gauge theories~\cite{Peskin,Zee}. A proposed chiral gauge theory must be anomaly free. Otherwise it is not renormalizable. Arguably the most important example of the anomaly-free rule is the Glashow-Salam-Weinberg electroweak theory, in which the 4 triangle anomalies linking the lepton and quark doublets with gauge bosons sum exactly to zero within each generation. This fortuitous cancellation has been called ``magical''~\cite{Peskin}.

In 1983, Nielsen and Ninomiya (NN) ~\cite{Nielsen} proposed that the chiral anomaly may be observable in a crystal with massless Dirac states in even spacetime dimension (1+1 or 3+1) (Fig. \ref{figWeyl}C). When we set the mass $m$= 0 in the Dirac equation, the Dirac 4-spinor $\Psi$ describes two independent populations of chiral particles with spin locked either parallel to their momentum (and described by the Weyl 2-spinor $\psi_R$) or antiparallel ($\psi_L$). The chiral charges are separately conserved. This chiral symmetry is ruined as soon as we couple the particles to $F_{\mu\nu}$ (by applying an $E$ field). Chiral charge is pumped from one branch to the other at a rate given by the triangle anomaly, viz.
\be
W = \chi\frac{e^3}{4\pi^2\hbar^2} {\bf E\cdot B}.
\label{eq:W}
\ee
At first blush, this would appear to be a restatement of charge transport in an ordinary metal. If we regard the left and right-moving states as opposite limbs of a large Fermi surface (FS), the $E$ field simply biases the occupation of the two limbs leading to the familiar FS shift in $k$-space. However, as discussed below, the implied ``locking'' of $\bf E$ to $\bf B$ is unusual and new. It actually leads to features that fall beyond the purview of standard transport in metals.

\begin{figure}[t]
\includegraphics[width=9cm]{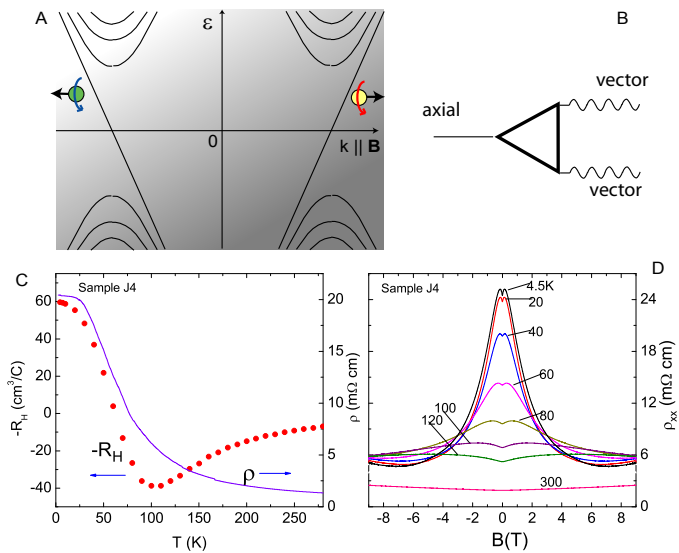}
\caption{\label{figRho} 
Panel A: Sketch of the Landau levels (LL) in a Weyl semimetal showing chiral states in the lowest LL with opposite velocities and chiralities (arrows) $\parallel\bf B$. An $\bf E$-field $\parallel \bf B$ breaks chiral symmetry and leads to an axial current. 
Panel B shows the triangle anomaly that ruins the conservation of chiral charge. 
Panel C: The $T$ dependence of the resistivity $\rho$ in $B=0$ and Hall coefficient $R_H$ in Na$_3$Bi.
$R_H$ is measured in $B<$2 T applied $\parallel \bf c$. At 3 K, $R_H$ corresponds to a density $n = 1.04\times 10^{17}$ cm$^{-3}$. The inset shows the contact labels and the $x$ and $y$ axes fixed to the sample. 
Panel D: Curves of the longitudinal magnetoresistance $\rho_{xx}(B,T)$ at selected $T$ from 4.5 to 300 K measured with $\bf B\parallel \hat{x}$ and $I$ applied to (1,4). The steep decrease in $\rho_{xx}(B,T)$ at 4.5 K reflects the onset of the axial current in the lowest LL. Adapted from Xiong \etal~\cite{Xiong2}.
}
\end{figure}

Unsurprisingly, the discussion in 2011 of Weyl nodes in the context of 3D topological phases of matter~\cite{Ashvin,Balents,Ran,Son,Sid,Hosur} stimulated intense renewed interest in NN's proposal, initially among theorists but now extending to experimentalists as well. Theoretically, the states in the two Weyl nodes are quantized into Landau levels in strong $B$. A key feature of the Weyl spectrum is that the $n$ = 0 Landau level (LL) is chiral, with velocity $\bf v$ parallel or antiparallel to $\bf B$ (Fig. \ref{figRho}A). We have precisely the situation described by NN, except that the 1D dispersion direction is controlled by $\bf B$.

\begin{figure}
\includegraphics[width=10cm]{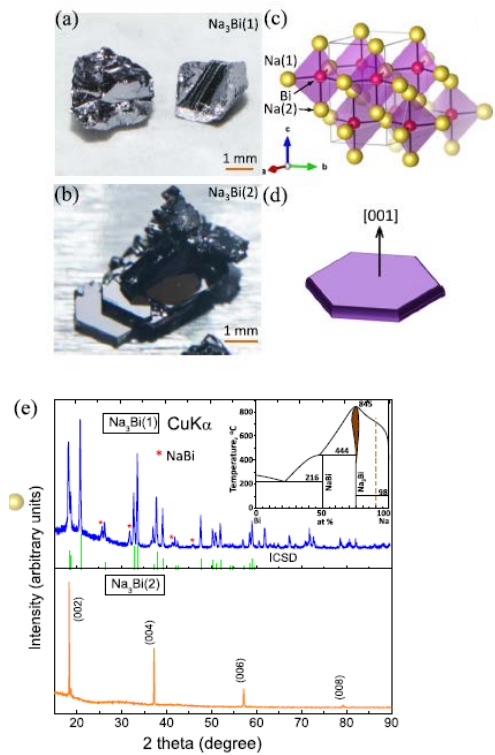}
\caption{\label{figxtal} 
Crystal growth and structure analysis. (a) Photograph of stoichiometric cooled Na$_3$Bi (1) crystals with naturally
cleaved pristine faces. (b) Photograph of typical 90$\%$ Na-flux grown Na$_3$Bi (2) crystals. (c) The
honeycomb crystal structure of Na$_3$Bi. The Na(2) atoms are cross linked to Bi of -Na(1)-Bi-Na(1)-Bi hexagonal layers. (d) 
Sketch of flux-grown crystals showing the axis [001]. (e) Recorded X-ray diffractograms. The upper trace (blue) is for powder specimen of Na$_3$Bi (1) crystals and the lower trace (orange) is for Na$_3$Bi (2) single crystal. The inset in (e) shows a schematic of the Na-Bi phase diagram. The vertical dashed line indicates the Na-rich composition used in the flux growth. Adapted from Kushwaha \etal~\cite{Satya}.
}
\end{figure}

As mentioned, the breaking of time-reversal symmetry (TRS) by $\bf B$ causes each Dirac node to split into two Weyl nodes~\cite{Bernevig,Wang1,Wang2}. In the absence of an electric field $\bf E$, the two chiral populations are separately conserved (Fig. \ref{figWeyl}A). However, application of $\bf E$ ruins the conservation and leads to a charge-pumping process between nodes, corresponding to the chiral anomaly sketched in Fig. \ref{figWeyl}B~\cite{Nielsen,Ashvin,Balents,Son,Sid,Hosur}. As shown in Fig. \ref{figWeyl}A, the lowest Landau level (LL) of the Weyl states in a crystal is chiral. For $\chi = 1$ the lowest LL is right-moving (velocity $\bf v\parallel \bf {B}$), whereas for $\chi=-1$ it is left-moving.~\cite{Nielsen,Ashvin,Balents,Son,Hosur} At steady state, $\bf E$ transfers charge from say the $\chi = -1$ branch to the $\chi= 1$ branch at the rate~\cite{Hosur}

The rate peaks for $\bf E\parallel B$ and vanishes when $\bf E\perp B$. In the weak-$B$ limit, a Berry curvature approach gives the chiral conductivity~\cite{Son} 
\be
\sigma_{\chi} = \frac{e^2}{4\pi^2\hbar c}\frac{v}{c}\frac{(eBv)^2}{\epsilon_F^2}\tau_v,
\label{eq:SS}
\ee
where $\tau_v$ is the intervalley life-time and $\epsilon_F$ the Fermi energy.

\section{Dirac semimetal Na3Bi}
The Dirac semimetal Na$_3$Bi grows as mm-sized, deep-purple, plate-like crystals with the largest face parallel to the $a$-$b$ plane ($\bf \hat{c}$ is normal to the planes) (Fig. \ref{figxtal}a,b). We annealed the crystals for 10 weeks before opening the growth tube. Details of the growth and characterization are reported in Kushwaha \etal~\cite{Satya}. To avoid oxidation, crystals were contacted using silver epoxy  in an Argon glove box, and then immersed in paratone in a capsule before rapid cooling. In Na$_3$Bi, the Dirac nodes are located at the wave vectors $(0,0, \pm k_D)$ with $k_D \simeq 0.1\; \mathrm{\AA}^{-1}$~\cite{Chen,Suyang}. Initial experiments in our lab~\cite{Xiong1} on samples with a large Fermi energy $\epsilon_F$ (400 mV) showed only a positive MR with the anomalous $B$-linear profile reported in Cd$_3$As$_2$~\cite{Liang}.

Recent progress in lowering $\epsilon_F$ has resulted in samples that display a non-metallic resistivity $\rho$ vs. $T$ profile and a low Hall density $n\sim 1\times 10^{17}$ cm$^{-3}$ (Fig. \ref{figRho}C). We estimate the Fermi wavevector $k_F = 0.012 \;\rm{\AA}^{-1}$ ($8\times$ smaller than $k_D$). The unusual profiles of $\rho$ and the Hall coefficient $R_H$ in Panel C imply the zero-$B$ energy spectrum shown in Panel B. Below $\sim$10 K, the conductivity is largely due to electrons in the conduction band with electron mobility $\mu\sim$ 2,600 cm$^2$/Vs). Because the energy gap is zero, holes in the valence band are copiously excited even at low $T$. As $T$ rises above 10 K, the increased hole population leads to a steep decrease in $\rho$ and an inversion of the sign of $R_H$ at 62 K. From the maximum in $R_H$ at 105 K, we estimate that $\epsilon_F\sim 3k_BT\sim$ 30 mV. As shown in Fig. \ref{figRho}D, the resistivity $\rho_{xx}$ in a longitudinal field ($\bf B||I $, the current) displays a remarkable peak at 4.5 K corresponding to a large negative MR (the resistance measured is $R_{14,23}$ ($I$ applied to contacts 1 and 4, and voltage measured between contacts 2 and 3; the inset in Fig. \ref{figWeyl}C shows the contact labels and the $x$ and $y$-axes). Raising $T$ above $\sim$100 K suppresses the peak. The small density $n$ implies that $\epsilon_F$ enters the lowest ($N=0$) LL at $B\sim$ 4-6 T.

\section{A narrow current plume}
The axial current is predicted to be large when $\bf B$ is aligned with $\bf E$. A valuable test then is the demonstration that, if $\bf E$ is rotated by 90$^\circ$, the negative magnetoresistance (MR) pattern rotates accordingly, i.e. the axial current maximum is selected by $\bf B$ and $\bf E$, rather than being pinned to a crystal axis, even in the weak-$B$ regime.

\begin{figure}
\includegraphics[width=9 cm]{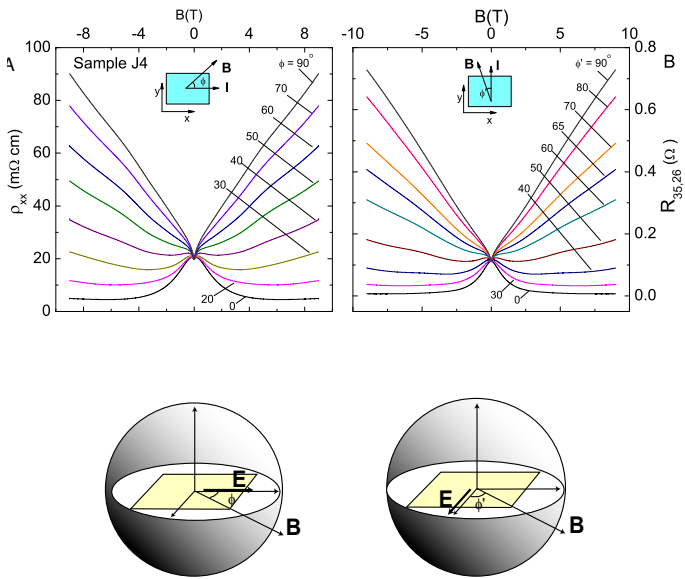}
\caption{\label{figMRXY}
Evidence for axial current in Na$_3$Bi (J4) obtained from transport measurements in an in-plane field $\bf B$. Panel A shows plots of the resistivity $\rho_{xx}$ vs. $B$ at selected field angles $\phi$ to the $x$-axis (inferred from resistance $R_{14,23}$, see inset). For $\phi$ = 90$^\circ$, $\rho_{xx}$ displays a $B$-linear positive MR. However, as $\phi\to 0^\circ$ ($\bf B||\hat{x}$), $\rho_{xx}$ is strongly suppressed. Panel B shows plots of $R_{35,26}$ with $\bf E$ rotated by 90$^\circ$ relative to Panel A ($\bf B$ makes angle $\phi'$ relative to $\bf\hat{y}$; see inset). The resistance $R_{35,26}$ changes from a positive MR to negative as $\phi'\to 0^\circ$. In both configurations, the negative MR appears only when $\bf B$ is aligned with $\bf E$. The spheres in the insets show the field orientations in 3D perspective. Adapted from Xiong \etal~\cite{Xiong2}.
}
\end{figure}

To test the anisotropy, we rotate $\bf B$ in the $x$-$y$ plane while still monitoring the resistance $R_{14,23}$. Figure \ref{figMRXY}A shows the curves of the resistivity $\rho_{xx}$ vs. $B$ measured at 4.5 K at selected $\phi$ (the angle between $\bf B$ and $\bf\hat{x}$). The MR is positive for $\phi = 90^\circ$ ($\bf B||\hat{y}$), displaying the nominal $B$-linear form observed in Cd$_3$As$_2$~\cite{Liang} and Na$_3$Bi~\cite{Xiong1} with $\bf B||c$. As $\bf B$ is rotated towards $\bf \hat{x}$ ($\phi$ decreased), the MR curves are pulled towards negative values. At alignment ($\phi = 0$), the longitudinal MR is very large and fully negative (see SI for the unsymmetrized curves as well as results from a second sample). 

We then repeat the experiment \emph{in situ} with $I$ applied to the contacts (3, 5), so that $\bf E$ is rotated by 90$^\circ$ (the measured resistance is $R_{35,26}$). Remarkably, the observed MR pattern is also rotated by 90$^\circ$, even when $B<$ 1 T. Defining the angle of $\bf B$ relative to $\bf\hat{y}$ as $\phi'$, we now find that the MR is fully negative when $\phi'$ = 0. The curves in Panels A and B are nominally similar, except $\phi = 0$ and $\phi'=0$ refer to $\bf\hat{x}$ and $\bf\hat{y}$, respectively. 

We remark on the two striking features in Figs. \ref{figMRXY}A and \ref{figMRXY}B. The appearance of a strongly varying in-plane MR is rare in metals. The closest example we are aware of is the ``anisotropic MR'' (AMR) observed in ferromagnets~\cite{McGuire,Campbell,Checkelsky}. In an in-plane $\bf B$, one observes $\rho(\phi) \sim P + Q\sin^2\phi$ caused by anisotropic scattering of carriers with velocity $\bf v||M$ vs. $\bf v\perp M$ ($\bf M$ is the magnetization and $P$ and $Q$ are positive parameters). However, as AMR always leads to a positive MR, it cannot account for the very large negative MR observed when $\phi$ (or $\phi'$) $\to 0$. Moreover, no magnetic transitions have been detected in Na$_3$Bi down to 2 K.

\begin{figure}[t]
\includegraphics[width=9 cm]{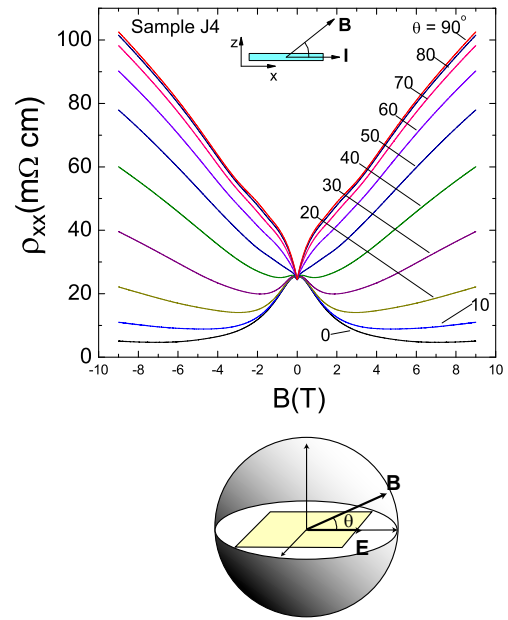}
\caption{\label{figMRvertical} 
Magnetoresistance of Na$_3$Bi ($R_{14, 23}$) when $\bf{B}$ lies in the $x$-$z$ plane at selected angles $\theta$, with $\bf{E}$ 
$\parallel\bf\hat{x}$ axis ($\theta$ is the angle between $\bf B$ and $\bf\hat{x}$). As in Fig. 6, an enhanced current (large negative MR) is observed when $\bf B$ approaches alignment with $\bf E$. When $\bf{B}$ is perpendicular to $\bf{E}$, the MR has a positive $B$-linear profile. The field orientations are shown in 3D perspective in the lower insert. 
From Jun Xiong (unpublished).
}
\end{figure}

\begin{figure}[t]
\includegraphics[width=8cm]{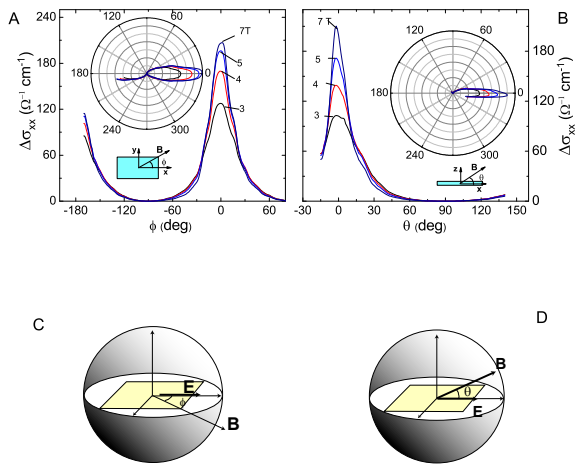}
\caption{\label{figPlume} 
Angular dependence of the axial current in Sample J4 at 4.5 K inferred from measurements of $R_{14,23}$ in tilted ${\bf B}(\theta,\phi)$. In Panel A, $\bf B$ lies in the $x$-$y$ plane at an angle $\phi$ to $\bf\hat{x}$ (sketch in inset). The 
conductance enhancement $\Delta \sigma_{xx}$ at fixed $B$ is plotted against $\phi$ for fields $3\le B\le 7$ T. The insets show the polar representation of $\Delta \sigma_{xx}$ vs. $\phi$. In Panel B, $\Delta \sigma_{xx}$ is plotted versus $\theta$ for $3\le B\le 7$ T for $\bf B$ lying in the $x$-$z$ plane. As sketched in the inset, $\theta$ is the angle between $\bf B$ and $\bf\hat{x}$. In Panels A (B), the axial current is peaked when $\phi\to 0$ ($\theta\to 0$) with an angular width that narrows significantly as $B$ increases. Panels C and D show the orientations of $\bf B$ and $\bf E$ relative to the crystal axes in Panels A and B, respectively. Adapted from Xiong \etal~\cite{Xiong2}.
}
\end{figure}

\section{Steering the plume by field}
The second feature is the tight linkage between the MR pattern and the common direction of $\bf B$ and $\bf E$ even in \emph{weak} $B$ ($\ll 1/\mu\sim$ 4 T). If the in-plane MR is caused by FS anisotropies (between $a$ and $b$ axes), the extrema of the oscillation ought to be pinned to the crystal axes. We should not be able to alter the resistivity tensor by rotating the weak $E$ and $B$ fields (this violates linear response). Viewed from ordinary transport, the observed rotation in weak $B$ is indeed anomalous.
However, it agrees with the prediction of the chiral anomaly -- the axial current peaks when $\bf E$ aligns with $\bf B$ even for weak fields.

The large negative MR in Panels A and B suggests a long relaxation time for the novel current.
We estimate the relaxation time $\tau_v$ for internode scattering from curves of the conductance $G = 1/R_{35,26}$ as follows. At low $B$, $G$ increases rapidly as $B^2$ consistent with Eq. \ref{eq:SS}. We form the ratio $\sigma_{\chi}/\sigma_0 = \frac34 (k_F\ell_B)^{-4}.(\tau_v/\tau_{tr})$ (where $\sigma_0$ is the Drude conductivity, $\ell_B$ is the magnetic length $\sqrt{\hbar/eB}$ and $\tau_{tr}$ the usual transport lifetime). Fitting to the observed parabolic curve, we find that $\tau_v/\tau_{tr}$ = 40-60. The scattering rate relaxing the axial current is anomalously low compared with the scattering rate $1/\tau_{tr}$ of the conventional states in zero $B$. 

\section{Angular width of plume}
A surprise to us is the acute sensitivity of the novel current to misalignment at large $B$. We have examined how the conductivity derived from $R_{14,23}$ decays as $\bf B$ is tilted away from $\bf\hat{x}$ in either the $x$-$y$ or the $x$-$z$ plane. Figure \ref{figPlume}A displays the curves of $\Delta\sigma_{xx}(B,\phi) = \sigma_{xx}(B,\phi)- \sigma_{xx}(B,90^\circ)$ vs. $\phi$ as $\bf B$ is tilted in the $x$-$y$ plane at an angle $\phi$ to $\bf\hat{x}$, with $B$ fixed at values 3$\to$7 T. Figure \ref{figPlume}B shows the same measurements but now with $\bf B$ lying in the $x$-$z$ plane at an angle $\theta$ to $\bf\hat{x}$. In both cases, the low-field curves ($B\le$ 2 T) are reasonably described with $\cos^p\phi$ (or $\cos^p\theta$) with $p$ = 4 (not shown). However, for $B>$ 2 T, the angular widths narrow significantly. Hence, at large $B$, the axial current is observed as a strongly collimated beam in the direction selected by $\bf B$ and $\bf E$ as $\phi$ or $\theta$ is varied.

To see what happens at larger $B$, we extended measurements of $R_{14,23}$ to $B$ = 35 T. 
We observe a new feature at $H_{k}\sim$ 23 T when $\bf B||\hat{y}$. As $\bf B$ is tilted away from $\bf\hat{y}$ ($\phi\to \; 55^\circ$), the feature at $H_k$ becomes better resolved as a kink. The steep increase in $\rho_{xx}$ above $H_k$ suggests an electronic instability at large $B$. However, as we decrease $\phi$ below $45^\circ$, $H_k(\phi)$ moves rapidly to above 35 T. The negative MR curve at $\phi = 0$ remains unaffected by the instability up to 35 T (the small rising background is from a weak $B_z$ due to a slight misalignment). 

To us, the unusual locking of the negative MR pattern in Figs. \ref{figMRXY}A and \ref{figMRvertical}B to $\bf E$ and $\bf B$ in weak $B$ constitutes strong evidence for the axial current. The experiment confirms the $B^2$ behavior in Eq. \ref{eq:SS} and provides a measurement of the long internode scattering lifetime. However, the width of the collimated beam in the direction of $\bf B$ is much narrower than expected from the theory. 

In addition to the results here, several groups have also reported observing the chiral anomaly in other materials (primarily TaAs and ZrTe$_5$). The evidence is by and large restricted to the appearance of negative longitudinal MR often bracketed by large positive MR at lower and higher $B$. A concern is that a negative, longitudinal MR restricted to a narrow field interval is (by itself) a rather slender reed to hang a weighty claim from. Further tests, such as the demonstration of the field-steering effect, would appear to be necessary. Nonetheless, the dramatic increase in experimental activity on a growing list of candidate Weyl semimetals is an encouraging sign for the field. We anticipate exciting experimental developments in the next few years.



\begin{thebibliography}{99}



\bibitem{Nielsen} H. B. Nielsen and M. Ninomiya,
``The Adler-Bell-Jackiw Anomaly and Weyl Fermions in a Crystal,''
Phys Lett B {\bf 130}, 389-396 (1983).


\bibitem{Ashvin}  X. G. Wan, A. M. Turner, A. Vishwanath and S. Y. Savrasov,
``Topological semimetal and Fermi-arc surface states in the electronic structure of pyrochlore iridates,'' Phys Rev B {\bf 83}, 205101 (2011).


\bibitem{Balents} A. A. Burkov, M. D. Hook and L. Balents,
``Topological nodal semimetals,'' 
Phys Rev B {\bf 84}, 235126 (2011).

\bibitem{Ran} Kai-Yu Yang, Yuan-Ming Lu, and Ying Ran, 
``Quantum Hall effects in a Weyl semimetal: Possible application in pyrochlore iridates,''
Phys Rev B {\bf 84}, 075129 (2011).

\bibitem{Hosur} Pavan Hosur and Xiaoliang Qi 
``Recent developments in transport phenomena in Weyl semimetals,'' 
Comptes Rendus Physique {\bf 14}, 857-870 (2013), doi:10.1016/j.crhy.2013.10.010.

\bibitem{Son} D. T. Son and B. Z. Spivak,
``Chiral anomaly and classical negative magnetoresistance of Weyl metals,'' 
Phys Rev B {\bf 88}, 104412 (2013).



\bibitem{Sid} S.A. Parameswaran, T. Grover, D. A. Abanin, D. A. Pesin, A. Vishwanath,
``Probing the Chiral Anomaly with Nonlocal Transport in Three-Dimensional Topological Semimetals",
Phys. Rev. X  {\bf 4}, 031035 (2014).


\bibitem{Adler} Stephen L. Adler, 
``Axial-Vector Vertex in Spinor Electrodynamics,''
Phys. Rev. {\bf 177}, 2426 (1969).

\bibitem{Bell} J.S. Bell and R. Jackiw, 
Nuovo Cimento {\bf 60A}, 4 (1969).


\bibitem{Peskin} \emph{Introduction to Quantum Field Theory}, Michael E. Peskin and Dan V. Schroeder, 
(Westview Press, 1995), Ch. 19.

\bibitem{Zee} \emph{Quantum Field Theory in a Nutshell}, 2nd ed., A. Zee, (Princeton University Press, 2010), Part IV.7.

\bibitem{Murakami1} S. Murakami, 
``Phase transition between the quantum spin Hall and insulator phases in 3D: emergence of a topological gapless phase,'' 
New J. Phys. {\bf 9}, 356 (2007).

\bibitem{Murakami2} S. Murakami and S. Kuga, 
``Universal phase diagrams for the quantum spin Hall systems,''
Phys. Rev. B {\bf 78}, 165313 (2008).


\bibitem{Young} S. M. Young, S. Zaheer, J. C. Y. Teo, C. L. Kane, E. J. Mele and A. M. Rappe,
``Dirac Semimetal in Three Dimensions,'' 
Phys Rev Lett {\bf 108}, 140405 (2012).

\bibitem{Bernevig} Chen Fang, Matthew J. Gilbert, Xi Dai, and B. Andrei Bernevig,
``Multi-Weyl Topological Semimetals Stabilized by Point Group Symmetry,''
Phys Rev Lett {\bf 108}, 266802 (2012).


\bibitem{Wang1} Z. J. Wang, Y. Sun, X. Q. Chen, C. Franchini, G. Xu, H. M. Weng, X. Dai and Z. Fang,
``Dirac semimetal and topological phase transitions in $A_3$Bi ($A$ = Na, K, Rb),'' 
Phys Rev B {\bf 85}, 195320 (2012).

\bibitem{Wang2} Z. J. Wang, H. M. Weng, Q. S. Wu, X. Dai and Z. Fang,
``Three-dimensional Dirac semimetal and quantum transport in Cd$_3$As$_2$,'' 
Phys Rev B {\bf 88}, 125427 (2013).

\bibitem{Nagaosa} Bohm-Jung Yang and Naoto Nagaosa,
``Classification of stable three-dimensional Dirac semimetals with nontrivial topology,''
arXiv:1404.0754v1.



\bibitem{Chen} Z. K. Liu, B. Zhou, Y. Zhang, Z. J. Wang, H. M. Weng, D. Prabhakaran, S. K. Mo, Z. X. Shen, Z. Fang, X. Dai, Z. Hussain and Y. L. Chen,
``Discovery of a Three-Dimensional Topological Dirac Semimetal, Na$_3$Bi,''
Science {\bf 343}, 864-867 (2014).

\bibitem{Suyang} S.-Y. Xu, C. Liu, S. K. Kushwaha, T.-R. Chang, J. W. Krizan, R. Sankar, C. M. Polley, J. Adell, T. Balasubramanian, K. Miyamoto, N. Alidoust, G. Bian, M. Neupane, I. Belopolski, H.-T. Jeng, C.-Y. Huang, W.-F. Tsai, H. Lin, F. C. Chou, T. Okuda, A. Bansil, R. J. Cava, M. Z. Hasan, "Observation of a bulk 3D Dirac multiplet, Lifshitz transition, and nestled spin states in Na$_3$Bi", arXiv:1312.7624 (2013)


\bibitem{Neupane}Madhab Neupane, Su-Yang Xu, Raman Sankar, Nasser Alidoust, Guang Bian, Chang Liu, Ilya Belopolski, Tay-Rong Chang, Horng-Tay Jeng, Hsin Lin, Arun Bansil, Fangcheng Chou and M. Zahid Hasan,
``Observation of a three-dimensional topological Dirac semimetal phase in high-mobility Cd$_3$As$_2$
Nat. Commun. 5:3786 doi: 10.1038/ncomms4786 (2014).


\bibitem{Liu} Z. K. Liu, J. Jiang,	B. Zhou, Z. J. Wang, Y. Zhang, H. M. Weng, D. Prabhakaran, S-K. Mo, H. Peng, P. Dudin, T. Kim, M. Hoesch, Z. Fang, X. Dai, Z. X. Shen, D. L. Feng, Z. Hussain	, and Y. L. Chen, 
"A stable three-dimensional topological Dirac semimetal Cd$_3$As$_2$", 
Nat. Mater. {\bf 13}, 677-681 (2014).

\bibitem{Borisenko} Sergey Borisenko, Quinn Gibson, Danil Evtushinsky, Volodymyr Zabolotnyy, Bernd B\"{u}hner, and Robert J. Cava, 
"Experimental Realization of a Three-Dimensional Dirac Semimetal", 
Phys. Rev. Lett. {\bf 113}, 027603 (2014).


\bibitem{Yazdani} Sangjun Jeon, Brian B. Zhou, Andras Gyenis, Benjamin E. Feldman, Itamar Kimchi, Andrew C. Potter, Quinn D. Gibson, Robert J. Cava, Ashvin Vishwanath, and Ali Yazdani, 
"Landau quantization and quasiparticle interference in the three-dimensional Dirac semimetal Cd$_3$As$_2$,'' 
Nat. Mater. {\bf 13}, 851-856 (2014).

\bibitem{Liang} Tian Liang, Quinn Gibson, Mazhar N. Ali, Minhao Liu, R. J. Cava and N. P. Ong, 
``Ultrahigh mobility and giant magnetoresistance in the Dirac semimetal Cd$_3$As$_2$,''
Nat. Mater. (2014), DOI: 10.1038/NMAT4143

\bibitem{Xiong1} Jun Xiong, Satya K. Kushwaha, Jason Krizan, Tian Liang, R. J. Cava, and N. P. Ong,
``Anomalous conductivity tensor in the Dirac semimetal Na$_3$Bi,''
arXiv:1502.06266v1

\bibitem{Xiong2} Jun Xiong, Satya K. Kushwaha, Tian Liang, Jason Krizan, Wudi Wang, R. J. Cava, and N. P. Ong,
``Evidence for the chiral anomaly in the Dirac semimetal Na$_3$Bi,''
\emph{submitted}, Feb. 2nd, 2015.


\bibitem{Satya} 
Satya K. Kushwaha, Jason W. Krizan, Benjamin E. Feldman, András Gyenis, Mallika T. Randeria, Jun Xiong,
Su-Yang Xu, Nasser Alidoust, Ilya Belopolski, Tian Liang, M. Zahid Hasan, N. P. Ong, A. Yazdani, and R. J.
Cava, 
``Bulk crystal growth and electronic characterization of the 3D Dirac semimetal Na$_3$Bi,''
APL Materials {\bf 3}, 041504 (2015) [http://dx.doi.org/10.1063/1.4908158].

\bibitem{HasanFS} S.-Y. Xu, C. Liu, S. K. Kushwaha, R. Sankar, J. W. Krizan, I. Belopolski, M. Neupane, G. Bian, N. Alidoust, T.-R. Chang, H.-T. Jeng, C.-Y. Huang, W.-F. Tsai, H. Lin, P. P. Shivayev, F. C. Chou, R. J. Cava, M. Z. Hasan, “Observation of Fermi arc surface states in a topological metal,” Science {\bf 347}, 294-298 (2014).


\bibitem{McGuire} T. R. McGuire and R. I. Potter, IEEE Trans. Magn. {\bf 11}, 4 (1975).

\bibitem{Campbell} I. A. Campbell, A. Fert, and O. Jaoul, J. Phys. C {\bf 3}, S95 (1970).

\bibitem{Checkelsky} J. G. Checkelsky, Minhyea Lee, E. Morosan, R. J. Cava and N. P. Ong,
``Anomalous Hall effect and magnetoresistance in the layered ferromagnet Fe$_\frac14$TaS$_2$: The inelastic regime,''
Phys. Rev. B {\bf 77}, 014433 (2008).




\vspace{1cm}\noindent
$^*$Corresponding author. E-mail: npo@princeton.edu

\vspace{1cm}\noindent
{\bf Acknowledgments}\\
We acknowledge discussions with Andre Bernevig. The research is supported by the Army Research Office (ARO W911NF-11-1-0379) and by a MURI award for topological insulators (ARO W911NF-12-1-0461). The growth and characterization of crystals were performed by S.K, J.W.K. and R.J.C., with support from the National Science Foundation (NSF grant DMR 1420541). N.P.O. acknowledges the support of the Gordon and Betty Moore Foundation’s EPiQS Initiative through Grant GBMF4539. Some experiments were performed at the National High Magnetic Field Laboratory (NHMFL), which is supported by NSF Cooperative Agreement No. DMR-1157490, the State of Florida, and the U.S. Department of Energy. We thank Eun Sang Choi for assistance at NHMFL.















\end{thebibliography}
\end{document}